\documentstyle[aps,prl,twocolumn]{revtex}

\makeatletter
\renewcommand{\maketitle}{\@maketitle}
\makeatother

\newcommand{\ns}[1]{\hbox{$\!\!\!#1\!\!\!$}}

\def\inn{{\text{in}}}
\def\out{{\text{out}}}

\def\lb{{\hbox{[}}}
\def\rb{{\hbox{] }}}

\def\d{\displaystyle }
\begin{document}
\draft
\twocolumn[

\title{%
Asymptotic Theory  for the Probability Density Functions 
in Burgers Turbulence}
\author{Weinan E$^{a)}$ and Eric  Vanden Eijnden$^{b)}$}
\address{Courant Institute of Mathematical Sciences\\
New York University\\
New York, New York 10012}

\maketitle

\widetext
\leftskip 54.8pt
\rightskip 54.8pt
\begin{abstract}
A rigorous study is carried out for the  randomly forced Burgers equation
in the inviscid limit. No closure approximations
are made. Instead the probability density functions of
velocity and velocity  gradient are related to the statistics of
quantities defined along the shocks. This method  allows 
one to compute the anomalies, as well as asymptotics for
the structure functions and the probability density functions.
It is shown that the left tail for the probability density function
of the velocity gradient has to decay faster than $|\xi|^{-3}$. A further
argument confirms the prediction of E {\em et al.} \lb Phys. Rev. Lett. 
{\bf 78}, 1904 (1997)\rb that it should decay as $|\xi|^{-7/2}$.
\end{abstract} 

\pacs{PACS numbers: 47.27.Gs,  05.40.-a, 02.50.Ey}
]

\narrowtext

In this Letter, we focus on statistical properties
of solutions of the randomly forced Burgers equation
\begin{equation}
  \label{eq:burgers1}
  u_t+u u_x =\nu u_{xx} +f, 
\end{equation}
where $f$ is a zero-mean,  statistically homogeneous, 
white-in-time Gaussian process with covariance
\begin{equation}
  \label{eq:force}
  \langle f(x,t) f(y,s)\rangle = 2 B(x-y)\delta(t-s),
\end{equation}
where $B(x)$ is smooth.
We are particularly interested in  the probability density function (pdf)
of the velocity gradient $\xi(x,t)=u_x(x,t)$, 
since it depends heavily on the
intermittent events created by the 
shocks. Assuming statistical homogeneity, and 
letting $Q(\xi;t)$ be the pdf of $\xi(x,t)$, it can be 
shown that $Q$ satisfies
\begin{equation}
  \label{eq:Q1}
    Q_t=\xi Q +\bigl(\xi^2 Q\bigr)_\xi +B_1 Q_{\xi\xi}
      -\nu\bigl( \langle \xi_{xx}|\xi\rangle Q\bigr)_\xi,
\end{equation}
where  $B_1=-B_{xx}(0)$.
$\langle \xi_{xx}|\xi\rangle$ is the ensemble-average 
of $\xi_{xx}$ conditional on $\xi$. The explicit form of 
this term is unknown, leaving (\ref{eq:Q1}) unclosed.
There have been several proposals on how to approximately evaluate
the  quantity
\begin{equation}
  \label{eq:F1}
    F(\xi;t)=  -\lim_{\nu\to0}
      \nu\bigl( \langle \xi_{xx}|\xi\rangle Q\bigr)_\xi.
\end{equation}
At steady state, they all lead to an asymptotic expression of the form
\begin{equation}
  \label{eq:sQ1}
  Q \sim \left\{
  \begin{array}{ll}
    {\d C_-|\xi|^{-\alpha}}& {\rm as} \ \ 
    \xi\to-\infty,\\[6pt]
    {\d C_+\xi^\beta {\rm e}^{-\xi^3/(3B_1)}\qquad}& {\rm as} \ \ 
    \xi\to+\infty,
  \end{array}\right.
\end{equation}
for $Q$,
but with a variety of  values for the exponents $\alpha$ and $\beta$
(here the $C_\pm$'s are numerical constants).
By invoking the operator product expansion,
Polyakov\cite{pol95} suggested that $F=aQ+b\xi Q$, with $a=0$ and  $b=-1/2$.
This leads to $\alpha=5/2$ and $\beta=1/2$. Boldyrev\cite{bol97} 
considered the same closure
with $-1\leq b\leq 0$, which gives $2\leq \alpha\leq 3$ and $\beta=1+b$. 
The instanton analysis \cite{gumi96,bafako97}   
predicts the right tail of $Q$ without giving a precise value
for $\beta$, but has not given any specific prediction for the left tail.
 E~{\it et~al.}
\cite{ekhma97} made a geometrical
evaluation of the effect of $F$, based on the observation 
that large negative gradients are generated near shock creation. 
Their analysis
gives a rigorous upper-bound for $\alpha$: $\alpha \le 7/2$.
In~\cite{ekhma97}, it was claimed that this bound is actually 
reached, i.e., $\alpha=7/2$. 
Finally Gotoh and Kraichnan \cite{gokr98} argued that 
the viscous term is negligible to leading order for large ${|\xi|}$, i.e. 
$F\approx 0$ for  $|\xi|\gg B_1^{1/3}$. This approximation
leads to $\alpha=3$ and $\beta=1$. 
For other approaches, see e.g. \cite{bomepa95,bome96}.
In this letter we proceed at an exact evaluation of (\ref{eq:F1})
and we prove that $\alpha$ has to be strictly larger than $3$
(a result which does not require that steady state be reached). 
At steady state, we prove that $\beta=1$ and we give an argument 
which supports strongly the prediction of \cite{ekhma97}, namely,
$\alpha = 7/2$.

To begin with, let us remark that it is established in the
mathematics literature that the inviscid limit
\begin{equation}
  \label{eq:limit1}
  u^0(x, t)=\lim_{\nu \rightarrow 0} u(x, t),
\end{equation}
exists for almost all $(x,t)$. Since $u^0$ will in general develop
shocks, say, at $x=y$,  we may have $u^0_x 
\propto \delta(x-y)$, and 
one cannot simply drop the viscous term in the Burgers equation without giving
some meaning to $u^0 u^0_x$ at shocks.
This can be done using {\em BV-calculus}\cite{vol67}, which allows 
one to write an equation for $u^0$ and 
gives rules for manipulating the terms entering this equation and computing
the effect of the viscous term in the inviscid limit.
An alternative, more intuitive, way of accessing the effect of the viscous
shock on the velocity profile outside the shock is to carry out an
asymptotic analysis near and inside the shock. Here we will take the
second approach and refer the interested reader to \cite{eva99} for the
first approach with BV-calculus.
It is important to remark that the two approaches lead to the same
results.

Before considering velocity gradient, 
it is helpful to study the statistics of velocity itself.
Let $R(u;t)$ be the pdf of $u(x,t)$. Assuming statistical homogeneity,
$R$ satisfies
\begin{equation}
  \label{eq:R1}
    R_t=B_0 R_{uu}
    -\nu\bigl( \langle u_{xx}|u\rangle R\bigr)_u,
\end{equation}
where $B_0=B(0)$.
To compute  $-\nu\bigl( \langle u_{xx}|u\rangle R\bigr)_u$, 
let us note that for $\nu\ll 1$, the solutions of (\ref{eq:burgers1})
consist of 
smooth pieces where the viscous effect is negligible, separated
by thin shock layers inside which the viscous effect is important.
Let $u_\out(x,t)$ be the solution of the Burgers 
equation outside the viscous shock layer;
$u_\out$ can be obtained as a series expansion in $\nu$. To leading order
in $\nu$, $u_\out$ satisfies Riemann's equation, $u_t+uu_x=f$.
In order to deal with the shock layer, say at $x=y$, define 
\begin{equation} 
  \label{eq:uin1}
  u_\inn(x,t)=v\left(\frac{x-y}{\nu},t\right),
\end{equation}
and write $v= v_0 +\nu v_1+O(\nu^2)$.
To leading order, $v_0(z,t)$ satisfies 
$(v_0-\bar u){v_0}_z={v_0}_{zz}$,
yielding $v_0(z,t)=\bar u -(s/2)\tanh(sz/4)$ where
$\bar u =dy/dt$ and $s$ is the jump across the shock. Consequently we have 
the following generic velocity profile inside the shock layer:
\begin{equation}
  \label{eq:uin2}
  u_\inn(x,t)= \bar u-\frac{s}{2} \tanh \left(\frac{s (x-y)}{4\nu}\right)
  +O(\nu).
\end{equation}
The actual values of $\bar u$ and $s$ are obtained from
the matching conditions between $u_\inn$ and $u_\out$. In terms of $v$ and the 
stretched variable $z$, they are
\begin{equation}
  \label{eq:bc1}
  \lim_{z\rightarrow \pm\infty} v_0
  = \lim_{x- y\to0^\pm} u_\out= \bar u\pm\frac{s}{2}.
\end{equation}
It is well-known that $s \le 0$.

We will use (\ref{eq:uin2}) to evaluate the viscous term in (\ref{eq:R1}).
By definition \cite{rem0}, 
\begin{equation}
  \label{eq:uvisco1}
  \nu  \langle u_{xx} |u\rangle R =
  \nu\lim_{L\to\infty}\frac{1}{2L} \int_{-L}^{L} \!\!dx\ u_{xx}\,
  \delta[u-u(x,t)].
\end{equation}
In the limit $\nu\to0$ only small intervals around the shocks 
will contribute to the integral. So, we can split the integral into small 
pieces involving only the shock layers and use the generic form of $u_\inn$ 
in the layers to evaluate these integrals. To $O(\nu)$, this gives
\begin{equation}
  \label{eq:uvisco2}
  \begin{array}{l}
    {\d \nu   \langle u_{xx} |u\rangle R}\\[4pt]
    {\d=\nu\lim_{L\to\infty}\frac{N}{2L} \frac{1}{N}
      \sum_{j} \int_{\rm j-th\ layer}\hskip-5mm  dx\
      {u_\inn}_{xx}\, \delta[u-u_\inn(x,t)]}\\[8pt]
    {\d=\rho
      \int ds d\bar u \ T(\bar u,s;t)\int_{-\infty}^{+\infty}\!\! dz
      \ {v_0}_{zz}\, \delta[u-v_0(z,t)],}
  \end{array}
\end{equation}
where in the second integral we picked any particular shock layer
and we went to the stretched variable
$z=(x-y)/\nu$. Here $N$ denotes the number of shocks in $[-L,L]$,
 $\rho=\rho(t)=\lim_{L\to\infty}N/2L$ is the shock density, and 
$T(\bar u,s;t)$ is the probability density of $\bar u(y,t)$ and $s(y,t)$ 
conditional on the property that there is a shock at position $y$ ($T$
is independent of $y$ because of  statistical homogeneity).
The last integral in (\ref{eq:uvisco2}) can of course be evaluated using  
the explicit form of $v_0$.
Another, more elegant, way to proceed is to use the equation for $v_0$, 
$(v_0-\bar u){v_0}_z ={v_0}_{zz}$, and change the
integration variable from $z$ to $v_0$ using
$dz {v_0}_{zz}=dv_0{v_0}_{zz}/{v_0}_z= dv_0 (v_0-\bar u)$. The result is
\begin{equation}
  \label{eq:uvisco4}
   \lim_{\nu\to0}\nu \langle u_{xx} |u\rangle R =
     -\rho\int d s  \int_{u+s/2}^{u-s/2}\!\! d\bar u \ (u-\bar u)
     T(\bar u,s;t).
\end{equation}
This equation gives an {\em exact} expression for the viscous 
contribution in the limit $\nu \to 0$ in terms of
certain statistical quantities associated with the shocks.
Of course, using (\ref{eq:uvisco4}) in (\ref{eq:R1}) does not lead to a closed
equation since $T$ remains to be specified. However, information can already 
be obtained at this point without resorting to any closure assumption. 
For instance,  using (\ref{eq:uvisco4}) in 
(\ref{eq:R1}) and  taking the second moment of the resulting equation
yields $\langle u^2\rangle_t=2B_0-2\epsilon$ with
\begin{equation}
  \label{eq:epsilondef}
  \epsilon = \lim_{\nu\to0} \nu\langle u_x^2\rangle = \frac{1}{12}
  \rho \langle |s|^3\rangle.
\end{equation}
In particular, at steady state $\rho \langle |s|^3\rangle = 12 B_0$.

Similar calculations can be carried out for multi-point
pdf's and, in particular, for $W(w;x,t)$,  the 
pdf of the velocity difference $w (x,z,t) 
=u(x+z,t)-u(z,t)$. It leads to an equation of the form
\begin{equation}
  \label{eq:W1}
  \begin{array}{rcl}
    {\d W_t}&\ns{=}&{\d 
      -w W_x-2\int_{-\infty}^{w}\!\! dw'\  W_x(w';x,t)}\\[4pt]
    &&{\d +2[B_0-B(x)] W_{ww}+H(w;x,t),}
    \end{array}
\end{equation}
where, to $O(x)$, $H$ is given by
\begin{equation}
  \label{eq:H1}
  \begin{array}{rcl}
    {H}&\ns{=}&{\d \rho  \bigl[w S(w;t)+ \langle s\rangle \delta(w)\bigr]}
    \\[4pt]
    &\ns{+}&{\d 2\rho  \int_{-\infty}^w\! dw' \  S(w';t)-2\rho\theta(w)+O(x).}
  \end{array}
\end{equation}
Here $\theta(w)$ is the Heaviside function and
$S(s;t)=\int d\bar u\, T(\bar u,s;t)$ is the conditional pdf of 
$s(y,t)$. By direct substitution it may be
shown that the solution of (\ref{eq:W1}) is, to $O(x^2)$,\cite{rem1}
\begin{equation}
  \label{eq:sW1}
  W \sim (1-\rho x) \frac 1x Q\left(\frac{w}{x};t\right) 
  + \rho x S(w;t) + O(x^2).
\end{equation} 
The first term in this expression contains $Q(\xi;t)$, the
pdf of the non-singular part of the velocity gradient, to be considered below
(see (\ref{eq:nonsing})). This term accounts for those realizations of
the flow 
where there is no shock in between $z$ and $x+z$ (an event of probability
$1-\rho x+O(x^2)$). This term also leads to the consistency constraint 
that $\lim_{x\to 0} W = \delta(w)$ (using 
$\lim_{x\to 0} Q(w/x;t)/x= \delta(w)$). The next term in (\ref{eq:sW1}), 
$\rho x S(w;t)$, accounts for the realizations of the flow where there  
is a shock in between $z$ and $x+z$ (an event of probability
$\rho x+O(x^2)$).
Equation (\ref{eq:sW1}) can be used to compute the structure functions, 
$\langle |w|^a\rangle=\int dw\, |w|^a W$. To leading order this gives
\begin{equation}
  \label{eq:struct1}
  \langle |w|^a\rangle \sim
  \left\{
    \begin{array}{ll}
      {\d x^{a} \langle |\xi|^a\rangle+O(x)\quad}& \text {if}\ \ 0\le a<1,
      \\[5pt]
      {\d x \rho \langle |s|^a\rangle + O(x^{1+a})\quad}&\text {if}\ \ 1 < a,
    \end{array}\right.
\end{equation} 
where $\langle |\xi|^a\rangle=\int d\xi\, |\xi|^a Q$. Using 
$\rho \langle |s|^3\rangle =12 B_0$, we get Kolmogorov's relation for $a=3$
\begin{equation}
  \label{eq:kolmo1}
  \langle |w|^3\rangle \sim  12 x B_0.
\end{equation} 

We now go back to the velocity gradient.
Observe first that, in the limit $\nu\to0$, the velocity gradient
can be written as
\begin{equation}
  \label{eq:nonsing}
  u_x(x,t)=\xi(x,t)+\sum_j s(y_j) \delta(x-y_j),
\end{equation}
where the $y_j$'s are the locations of the shocks,
$\xi$ is the {\em non-singular} part of $u_x$.
Assuming homogeneity, a direct consequence of (\ref{eq:nonsing}) is 
\begin{equation}
  \label{eq:nonhomo}
  \langle u_x \rangle =
  \langle \xi\rangle + \rho \langle s\rangle=0.
\end{equation}
Unlike the viscous case where $\xi = u_x$, hence $\langle \xi \rangle =0$,
we have in the inviscid limit
  $ \langle \xi\rangle = -  \rho \langle s\rangle \not=0.$
Note also that the inviscid limit of the solutions of (\ref{eq:Q1})
converge to the pdf of $\xi$ only, which is still going to be denoted by  $Q$.

To evaluate $F$,
there are two ways to proceed. One is to rewrite (\ref{eq:W1}) in terms of the
pdf of $(u(x+z,t)-u(z,t))/x$ and take the limit as $x$ goes to zero.
This is the approach taken in~\cite{eva99}.
The other is to evaluate (\ref{eq:F1}) directly. The two approaches amount to 
different orders of taking the limit $ x \rightarrow 0, \nu \rightarrow 0$, and
give the same result. Hence the two limiting processes commute.
We will take the second approach and evaluate (\ref{eq:F1}) using the same 
basic idea as above. Here, however, we have to proceed
more carefully with the shock layer analysis. Differentiation of
(\ref{eq:uin2}) gives 
\begin{equation}
  \label{eq:xiin2}
  \xi_\inn(x,t)= -\frac{s^2}{8\nu} 
    {\rm sech}^2\left(\frac{s (x-y)}{4\nu}\right)+O(1).
\end{equation}
While the next order term in (\ref{eq:uin2}) was negligible in the limit
$\nu\to0$, the $O(1)$ contribution to $\xi_\inn(x,t)$
actually {\em dominates} the $O(\nu^{-1})$ contribution
at the border of the shock layer because the latter falls exponentially fast 
as the outer region is approached,
whereas the former tends to constants, say, $\xi_\pm$. In particular, the
matching between $\xi_\out(x,t)$ and $\xi_\inn(x,t)$  involves the 
$O(1)$ terms. To see how matching takes place,
differentiating the expression for $u_\inn$, we have
$\xi_\inn=\nu^{-1} {v_0}_z+{v_1}_z+O(\nu)$. The matching condition
between $\xi_\inn$ and $\xi_\out$ reads
\begin{equation}
  \label{eq:matchxi}
  \lim_{z\to\pm\infty}{v_1}_z=
    \lim_{x-y\to0^\pm}\xi_\out \equiv \xi_\pm.
\end{equation}
The equation for $v_1$ is
\begin{equation}
  \label{eq:burgersstretch1}
  {v_0}_t+(v_0-\bar u){v_1}_z+v_1 {v_0}_z ={v_1}_{zz}+f_x,
\end{equation}
and, from the above argument, the only information we really need 
about $v_1$ is its values at the boundaries $z\to\pm\infty$.
Since ${v_0}_z$ falls exponentially fast for large $|z|$,
(\ref{eq:burgersstretch1}) reduces to 
\begin{equation}
  \label{eq:burgersstretch1as}
  \bar u_t \pm \frac{s_t}{2}\pm \frac{s}{2} {v_1}_z= {v_1}_{zz}+f_x, 
  \qquad z\to\pm\infty,
\end{equation}
where we used the asymptotic values of $v_0$. Thus, as $z\to\pm\infty$,
\begin{equation}
  \label{eq:limv1}
  {v_1} \sim\mp\frac{2\bar u_t}{s}z-\frac{s_t}{s}z
  \pm\frac{2f_x}{s}z+c^\pm_1
  +c^\pm_2 {\rm e}^{\pm s z/2}.
\end{equation}
Notice that the exponential terms are irrelevant
in these expression since $ s \le 0$. Equation (\ref{eq:limv1}) implies
\begin{equation}
  \label{eq:limv1z}
  \lim_{z\to\pm\infty} {v_1}_z = \mp \frac{2\bar u_t}{s}-\frac{s_t}{s} 
  \pm\frac{2f_x}{s}
  =\xi_\pm,
\end{equation}
where the last equality is just  the definition of $\xi_\pm$.
 Note that (\ref{eq:limv1z})
can be  rewritten as
\begin{equation}
  \label{eq:limv2z}
  s_t = -\frac{s}{2}  \bigl( \xi_- +\xi_+\bigr), \qquad
  \bar u_t = \frac{s}{4}  \bigl( \xi_- -\xi_+\bigr)+f_x.
\end{equation}
In the limit $\nu\to0$ these are the  equations of motion along the
shock.

We can now evaluate the viscous contribution using
\begin{equation}
  \label{eq:diss1q}
  \nu \langle \xi_{xx} |\xi\rangle Q =
  \nu\lim_{L\to\infty}\frac{1}{2L} \int_{-L}^{L} dx\ \xi_{xx}\,
  \delta[\xi-\xi(x,t)].
\end{equation}
The calculation is similar to the one for the velocity and eventually leads to
\begin{equation}
  \label{eq:F2}
  F(\xi;t) =\frac{\rho}{2}
      \int d s \ s \bigl[V_-(\xi,s;t)+V_+(\xi,s;t)\bigr],
\end{equation}
where $V_\pm(\xi,s;t)$ are the conditional pdf's of 
$\xi_\pm(y,t)$ and $s(y,t)$. The appearance of $\xi_\pm$ in (\ref{eq:F2}) 
is of course a direct result of the $O(1)$ term in
(\ref{eq:xiin2}). 

We now use (\ref{eq:F2})
in  (\ref{eq:Q1}) and analyze some consequences of
\begin{equation}
  \label{eq:QQ1}
    Q_t=\xi Q +\bigl(\xi^2 Q\bigr)_\xi +B_1 Q_{\xi\xi}
    +F(\xi; t).
\end{equation}
Taking the first moment of (\ref{eq:QQ1}) leads to
\begin{equation}
  \label{eq:c21}
  \langle \xi\rangle_t=
  \left[\xi^3Q\right]_{-\infty}^{+\infty} + 
  \frac{\rho}{2} \left[\langle s\xi_-\rangle+\langle s\xi_+\rangle\right], 
\end{equation}
where we used $\int d\xi \ \xi F=
\rho[\langle s\xi_-\rangle+\langle s\xi_+\rangle]/2$.
On the other hand, averaging
the first equation in (\ref{eq:limv2z}) gives
\begin{equation}
  \label{eq:c23}
  \bigl(\rho\langle s\rangle\bigr)_t=
  -\frac{\rho}{2} 
  \left[\langle s\xi_-\rangle+\langle s\xi_+\rangle\right].
\end{equation}
This equation uses the fact that shocks are created at zero amplitude,
and shock strengths add up at collision.
These are consequences of the fact that the forcing is smooth in 
space \cite{rem2}.
Since $\langle \xi\rangle_t = -(\rho\langle s\rangle)_t$ from
(\ref{eq:nonhomo}), the comparison
between (\ref{eq:c21}) and (\ref{eq:c23}) 
tells us that the boundary term in (\ref{eq:c21}) must be zero. Since $Q\geq0$,
$\xi^3Q$ has different sign for large positive and large negative values of 
$\xi$. Therefore we must have
$\lim_{\xi\rightarrow+\infty}\xi^3 Q=0$ {\em and}
$\lim_{\xi\rightarrow-\infty}\xi^3 Q=0$. This proves that 
$Q$ goes to zero  {\em faster} than $|\xi|^{-3}$ as 
$\xi\rightarrow-\infty$ and $\xi\rightarrow+\infty$.

The analysis can be carried out one step further for the 
stationary case ($Q_t=0$). 
In this case, treating (\ref{eq:QQ1}) as an inhomogeneous second order ordinary
differential equation, we can write its general solution as
$Q = C_1 Q_1 +C_2 Q_2+Q_3$, where $C_1$ and $C_2$ are 
constants, $Q_1$ and $Q_2$ are  two linearly independent solutions
of the homogeneous equation associated with (\ref{eq:QQ1}), and
$Q_3$ is some particular solution of this equation.  One such particular 
solution is
\begin{equation}
  \label{eq:sQ2}
  Q_3=\int_{-\infty}^\xi \!\! d\xi'
      \ \frac{\xi' F(\xi')}{B_1}-\frac{\xi{\rm e}^{-\Lambda}}{B_1}
      \int_{-\infty}^\xi \!\! d\xi'
      \ {\rm e}^{\Lambda'} G(\xi'),
\end{equation}
where $\Lambda=\xi^3/(3B_1)$ and
\begin{equation}
  \label{eq:G1}
  G(\xi)=F(\xi)+\xi\int_{-\infty}^{\xi}\!\!d\xi'\ \frac{\xi' F(\xi')}{B_1}.
\end{equation}
With this particular solution,
it can be shown (see \cite{eva99} for details) 
that the realizability constraints imply that 
$C_1 = C_2 = 0$, i.e. the only non-negative, integrable solution is
$Q=Q_3$. Furthermore, in order
that $Q$ actually be non-negative, $F$ must satisfy
\begin{equation}
  {0\geq F \geq C\xi^2 {\rm e}^{-\xi^3/(3B_1)}\qquad {\rm as} \ \ 
    \xi\to+\infty,}
\end{equation}
for some constant $C<0$. 
Substituting into (\ref{eq:sQ2}), we get
\begin{equation}
  \label{eq:sQ3}
  Q \sim \left\{
  \begin{array}{ll}
    {C_- |\xi|^{-3}\int_{-\infty}^\xi d\xi'\ \xi' F(\xi')}& {\rm as} \ \ 
    \xi\to-\infty,\\[5pt]
    {\d C_+\xi {\rm e}^{-\xi^3/(3B_1)}\qquad}& {\rm as} \ \ 
    \xi\to+\infty,
  \end{array}\right.
\end{equation}
which confirms the result $Q\sim C_-|\xi|^{-\alpha}$ with $\alpha>3$ as 
$\xi\to-\infty$, and gives $\beta=1$.

The actual value of the exponent $\alpha$ depends 
on the asymptotic behavior of $F$. The latter can 
be obtained from further considerations on the dynamics of the shock
(\ref{eq:limv2z}).
This is rather involved and will be left to \cite{eva99}.
The result gives $\alpha=7/2$ which confirms the prediction of 
\cite{ekhma97}.
Here we will restrict ourselves to an interpretation of the
current approach in terms of the geometric picture.
Observe that the largest values of $\xi_\pm$ are achieved just after 
the shock formation. Assume that a shock is
created at time $t=0$, position $x=0$, and 
with velocity $u=0$. Then, locally
\begin{equation}
  \label{eq:shock1}
  x=ut-a u^3+\cdots.
\end{equation}
It follows that for $t\ll1$ the solutions of $0=ut-a u^3$, $u_{\pm}$,
behave as
\begin{equation}
  \label{eq:shock2}
  u_\pm=\mp \sqrt{\frac{t}{a}}\quad \Rightarrow
  \quad s= -2\sqrt{\frac{t}{a}},
\end{equation}
and ${\xi_\pm}$, solutions of $1=\xi t-3au^2\xi$, behave  as
\begin{equation}
  \label{eq:shock3}
  \xi_\pm = -\frac{1}{2t}.
\end{equation}
Assuming that these give the dominant contribution to $F(\xi)$ for large
negative values of $\xi$,  the asymptotic form of $F$ is
\begin{equation}
  \label{eq:Fasympt}
  F\sim C \int_0^\infty \!\! dt \  s(t) 
    \bigl\{\delta [\xi-\xi_-(t)]
   + \delta[\xi-\xi_+(t)]\bigr\},
\end{equation}
where $C$ is some constant 
related to the statistics of the shock life-time and $a$,
and $s(t)$, $\xi_\pm(t)$ are given by (\ref{eq:shock2}), (\ref{eq:shock3}). 
The evaluation of (\ref{eq:Fasympt}) 
gives $F\sim C |\xi|^{-5/2}$, and, hence,
\begin{equation}
  \label{eq:sQ4}
  Q \sim  C_- |\xi|^{-7/2}\qquad {\rm as} \ \  \xi\to-\infty.
\end{equation}
Even though this argument gives only a lower bound for $F$ at large
negative values of $\xi$, 
further arguments presented in \cite{eva99}
indicate that this lower bound is actually sharp.


We thank Bob Kraichnan and Stas Boldyrev for stimulating discussions.
The work of W. E is supported by a Presidential Faculty Fellowship from
the National Science Foundation.
The work of E. V. E. is supported by U.S. Department of Energy Grant
No. DE-FG02-86ER-53223.


\begin{references}
\bibitem{rem2} 
In cases (not considered here) when shocks are created at finite amplitude
(e.g. in the unforced case with piecewise linear initial data),
an additional term appears at the right hand side of (\ref{eq:c23}). 
Consequently from (\ref{eq:c21}), $Q$ 
decays like $|\xi|^{-3}$ when $\xi \to -\infty$.
\end{references}
\end{document}